\documentclass[english,aps,pra,twocolumn,superscriptadress]{revtex4}
\usepackage[T1]{fontenc}
\usepackage[latin9]{inputenc}
\usepackage{amstext}
\usepackage{graphicx}

\makeatletter
\@ifundefined{textcolor}{}
{%
 \definecolor{BLACK}{gray}{0}
 \definecolor{WHITE}{gray}{1}
 \definecolor{RED}{rgb}{1,0,0}
 \definecolor{GREEN}{rgb}{0,1,0}
 \definecolor{BLUE}{rgb}{0,0,1}
 \definecolor{CYAN}{cmyk}{1,0,0,0}
 \definecolor{MAGENTA}{cmyk}{0,1,0,0}
 \definecolor{YELLOW}{cmyk}{0,0,1,0}
 }

\makeatother

\usepackage{babel}
\begin{document}

\title{Tomography and state reconstruction with superconducting single-photon
detectors}

\author{J. J. Renema\textsuperscript{1)}, G. Frucci\textsuperscript{2)},
M.J.A. de Dood\textsuperscript{1)}, R. Gill\textsuperscript{3)},
A. Fiore\textsuperscript{2)}, M.P. van Exter\textsuperscript{1)}}

\affiliation{1) Leiden Institute of Physics, Leiden University, Niels Bohrweg
2, 2333 CA Leiden, the Netherlands }

\affiliation{2) COBRA Research Institute, Eindhoven University of Technology,
P.O. Box 513, 5600 MB Eindhoven, The Netherlands }

\affiliation{3) Mathematics Institute, Leiden University, Niels Bohrweg 1, 2333
CA Leiden, the Netherlands}
\begin{abstract}
We investigate the performance of a single-element superconducting
single-photon detector (SSPD) for quantum state reconstruction. We
perform quantum state reconstruction, using the measured photon counting
behavior of the detector. Standard quantum state reconstruction assumes
a linear response; this simple model fails for SSPDs, which are known
to show a non-linear response intrinsic to the detection mechanism.
We quantify the photon counting behaviour of the SSPD by a sparsity-based
detector tomography technique and use this to perform quantum state
reconstruction of both thermal and coherent states. We find that the
nonlinearities inherent in the detection process enhance the ability
of the detector to do state reconstruction compared to a linear detector
with similar efficiency for detecting single photons.
\end{abstract}
\maketitle

\section{Introduction}

We investigate the photon counting abilities of Superconducting Single-Photon
Detectors (SSPDs) \citep{Goltsman2001}, motivated by the recent upsurge
in the use of such detectors in quantum optics \citep{Zinoni2007,Natarajan2010}
and quantum cryptography \citep{Hadfield2006,Collins2007} experiments.
SSPDs are fast \citep{Kerman2006}, spectrally broadband \citep{Verevkin2004,Korneev2004,Verevkin2002}
single-photon sensitive detector with low noise \citep{Korneev2005}.
These detectors consist of an ultrathin meandering strip of a superconductor
with low Cooper pair density, typically NbN. When biased close to
the critical current, the absorption of a photon produces a transition
from the superconducting to the normal state, resulting in the creation
of a resistive area and the appearance of a voltage pulse in the external
read-out circuit.

It was previously shown \citep{Goltsman2001,Divochiy2008,Renema2012,Akhlaghi2009,Bitauld2010}
that depending on the bias current through the superconductor, the
detector has multiphoton regimes, where the energy from several photons
is required to break the superconductivity. These multiphoton detection
events - which depend on several photons being absorbed close together
- increase the probability of the detector clicking in a nontrivial
way \citep{Akhlaghi2009a,Akhlaghi2009,Renema2012}.

Quantum state reconstruction involves finding the photon number distribution
of an unknown quantum state of light from the response of some detector.
This task is of fundamental importance for any quantum optics or quantum
communication experiment, as the final step in such an experiment
is always the measurement of the photon occupation number in a particular
detection mode. Surprisingly, the reconstruction of a radiation field
can be one with a single detector that has only an on/off output \citep{Zambra2005}.
This is possible because measuring the count rate at different settings
of the detector (e.g. at different efficiencies) produces a \emph{response
curve} characteristic for each photon number. The reconstructed distribution
of photon numbers for an unknown state is then given by the linear
combination of response curves that best describes the measured count
rates of that state \citep{Mogilevtsev1998,Rossi2004,Zambra2005,Rehacek2003,Banaszek1998,Banaszek1999,Banaszek1996,Rossi2004a}.
By taking into account the finite efficiency of the detector, it is
possible to reconstruct the state at the input of the detector rather
than the statistics of the absorbed photons.

The results in this paper are divided into three parts. In Section
III, we perform a modified version of detector tomography \citep{Lundeen2008}
on the SSPD. This tomography quantifies the complex behaviour of the
device, enabling the use of the SSPD in situations where responses
to several different photon numbers are important. By using a technique
which has minimal assumptions, we overcome the problem that the understanding
of the working of the detector is still incomplete \citep{Hofherr2010},
harnessing the SSPD for quantitative multiphoton applications. Our
tomographic technique is based on sparsity. The advantage of a sparsity-based
technique is its robustness \citep{Renema2012}. We show explictly
how to apply this tomographic technique to an SSPD.

Next, in Section IV, we perform quantum state reconstruction with
the SSPD. We reconstruct states with average photon numbers up to
$\langle n\rangle=11.4$, thereby showing that the tomographic process
was succesful, and demonstrating that it is possible to use a nonlinear
device for quantum state reconstruction. We reconstruct both coherent
and thermal states.

Lastly, in Section V, we evaluate the effect of these nonlinearities
on the quality of the state reconstruction process. By evaluating
the Cramer-Rao bound - the theoretical limit on the amount of information
that may be extracted from a measurement - we establish that the intrinsic
nonlinearities of an SSPD are benificial for quantum state reconstruction.

\section{Theory: quantum state reconstruction}

The goal of Quantum State Reconstruction is to reconstruct the diagonal
elements $\rho_{nn}=\mathrm{\textrm{diag}}(\rho)$ of the input state
of the light. From observations at different configurations of the
detector, it is possible to reconstruct the state because each photon
number gives a particular response on the detector that is characteristic
of that photon number. The task is to find from the set of response
curves the linear combination of photon numbers that best describes
the measured count rate of some unknown state. The response of the
detector at each setting $\nu$ of the tuning parameter is described
by a Positive Operator-Valued Measure (POVM) element $\Pi_{\nu}=\Sigma_{n}\Pi_{\nu n}|n\rangle\langle n|$,
and the detector responds to the state with a probability $R_{\nu}=Tr(\rho\Pi_{\nu})$.
The POVM contains a full quantitative description of the measurement
process.

Due to the shot noise associated with the discreteness of the photon
counting process, the problem of solving this set of equations simulataneously
is inevitably statistical in nature, since the equations will not
be analyticaly invertible. This problem can be solved by a maximum
likelihood (ML) technique, using the Expectation Maximization (EM)
algorithm \citep{Mogilevtsev1998,Rossi2004,Zambra2005,Rehacek2003,Banaszek1998,Banaszek1999,Banaszek1996,Rossi2004a}
to find the best solution, while respecting the normalization of the
state. A derivation is given in \citep{Banaszek1998}. The i-th iteration
of this algorithm is given by:

\begin{equation}
\rho_{nn}^{(i+1)}=\rho_{nn}^{(i)}\sum_{\nu=1}^{N_{0}}\frac{\Pi_{\nu n}}{\sum_{\lambda}\Pi_{\lambda n}}\frac{R_{\nu}}{p_{\nu}(\rho^{(i)})},
\end{equation}
where $\rho^{(i)}$ is the state at the i-th iteration, $N_{0}$ is
the total number of experimental preparations, $R_{\nu}$ is the measured
click probability at the $\nu$-th experimental configuration and
$p_{\nu}(\rho^{(i)})$ is the calculated click probability at the
$\nu$-th experimental configuration. It is known that this algorithm
converges to the ML solution, for which the standard errors are given
by the Cramer-Rao bound \citep{Rossi2004a}. It is also known that
this algorithm can take many iterations to converge. Following earlier
work \citep{Mogilevtsev1998,Rossi2004,Zambra2005,Rehacek2003,Banaszek1998,Banaszek1999,Banaszek1996,Rossi2004a},
we take our number of iterations to be $10^{6}$.

\section{Experimental setup}

The SSPD used in this experiment is a commercial NbN meander produced
by Scontel. The width of the wire is 100 nm, and the distance between
the wires is 150 nm. The size of the active area is 10 $\mu$m by
10 $\mu$m. The device was cooled in a bath cryostat to a temperature
of 1.7 K. The measured overall system quantum efficiency for the one-photon
Fock state was 2.8\% at a bias current of 13.3 $\mu$A (corresponding
to $I_{b}/I_{c}\approx0.9$) and a wavelength of $\lambda=1500$ nm.

For our detector tomography procedure, we illuminate the device with
a series of coherent states varying from 130 fW to 108 nW (0.05 to
4.1$*10^{4}$ photons/pulse). The low powers were achieved with a
computer-controlled variable attenuator, whose linearity to -60 dB
was verified independently. From the measured response to coherent
states, we reconstruct the POVM using the method described below.
The coherent states were generated by a Fianium supercontinuum pulsed
laser. The repetition rate of this laser was 20 MHz, the specified
pulse width < 7 ps. The light was filtered to have a center wavelength
$\lambda_{0}=1500$ nm and a spectral width $\Delta\lambda=12$ nm.
The observed POVM was then used to reconstruct coherent and thermal
states. We verified independently that the output from our supercontinuum
laser is indeed a coherent state. We measure $g^{(2)}(0)$ with a
coincidence circuit and obtain $g^{(2)}(0)=0.97\pm0.02$.

We generate pseudothermal states by the standard technique of a rotating
ground glass plate \citep{Martienssen1964}, which was illuminated
with the coherent states described above. The exponential probability
distribution of the intensity of the resulting speckles creates photon
statistics that are equivalent to thermal light when averaged over
many realizations of the angle setting of the plate.

Unfortunately, after the reconstruction of the coherent states, the
alignment of the detector in the cryostat was degraded. We therefore
recharacterized the device in its new configuration with a set of
coherent states before performing the reconstruction of the thermal
states. The degradation manifests itself as an increased dark count
probability, which was 0.01 / pulse at $I_{b}/I_{c}\approx0.9$.

\section{Sparsity-based tomography}

We start by measuring the detector response curves, i.e. the detection
probablity versus detector bias current, for a set of coherent states.
From these, we deduce the more fundamental response curves for Fock
states. Fig. 1 shows the resulting set of inferred detector response
curves, i.e. the probability of the detector to respond to a certain
Fock state. The detector tomography is performed by a method based
on the one described in \citep{Renema2012}. The essential assumption
is one of sparsity: we describe the detector by as few physical parameters
as possible, while not restricting the possible range of behaviours
that our model describes. More specifically, we describe our detector
by a combination of linear attenuation (given by linear losses in
the detection process such as the finite absorption of the NbN layer)
followed by a nonlinear photodetection process inside the NbN layer.
The reason for including a linear efficiency seperately is that it
significantly reduces the number of parameters required to model the
detector, making the tomography more robust.

A second reason is that nonunity linear efficiency introduces correlations
between the various$\Pi_{\nu n}$ at one bias current. The reason
for this is that at nonunity efficiency, the $n$ photons necessary
for an $n$-photon nonlinear process could have come from any $N>n$
number of incident photons. By explicitly including this effect, we
make sure that our reconstructed POVM is compatible with this process.

This description in terms of a linear absorber and a nonlinear process,
which we showed to be applicable to the NbN nanodetector \citep{Bitauld2010}
is applicable to the SSPD as well. We can therefore write for the
click probability:

\begin{equation}
R_{\nu}=e^{-\eta\langle n\rangle}\sum_{k=0}^{k=\infty}p_{\nu k}\frac{(\eta\langle n\rangle)^{k}}{k!},
\end{equation}
where $\eta$ is the linear efficiency, $\langle n\rangle$ is the
mean photon number, $R_{\nu}$ is the count rate at a given bias current
and the $p_{\nu k}$ are the POVM elements prior to the inclusion
of the finite linear efficiency. These elements now only include the
nonlinear effects of the detector, i.e. the photon number threshold
regime that the detector is in, which depends on the bias current.

After each fit, where the fits at different currents are completely
independent, we reinclude the $\eta$ into the $p_{\nu k}$ to produce
the POVM element $\Pi_{\nu k}$ by the following procedure: first,
we fix the number $n_{mr}$ at which we are going to truncate the
Hilbert space for the reconstruction. Then, we construct a vector
of length $n_{mr}$, where the first 5 elements are $p_{\nu n},$
and the other elements are equal to 1. Finally, we multiply this vector
by a Bernouilli transformation \citep{Scully1969} $L_{kk'}=\left(\begin{array}{c}
k\\
k'
\end{array}\right)\eta^{k'}(1-\eta)^{k-k'}$ , absorbing the linear losses into the POVM. We perform state reconstruction
with the POVM consisting of all $\Pi_{\nu k}$ obtained at different
currents.

For this experiment, we are not interested in separating the linear
and nonlinear effects, but rather in finding a description of the
entire detector. Therefore, we truncate the sum at $n_{max}=4$ for
all currents. This is equivalent to assuming that the detector is
governed only by linear effects at sufficiently high photon numbers.
This choice is motivated by the fact that we do not enter the three-photon
regime in the current range over which we operate our detector and
is justified by the good fits obtained with this model.

For the analysis, we grid our measured count rates by linear interpolation,
producing 165 current settings, from 5 $\mu$A to 13.25 $\mu$A, which
is the current range over which we could measure count rates at enough
powers to create a good fit. This is also the current range over which
we perform the state reconstruction.

\begin{figure}
\includegraphics[width=10cm]{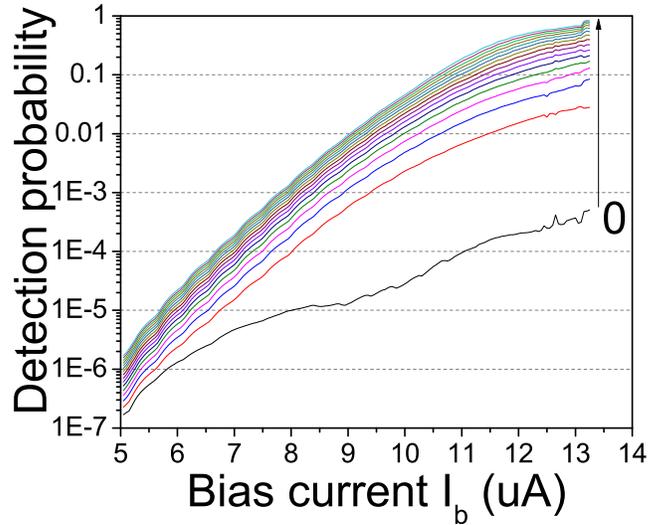}

\caption{Response curves inferred from detector tomography as a function of
bias current through the device. On the x axis is the bias current,
on the y axis is the probability that the detector responds to a particular
number of photons (Fock state). The black line indicates 0 photons,
the arrow indicates the direction of increasing photon number. Note
that we have shown only the first incident 15 photon numbers for clarity. }
\end{figure}

\section{Quantum state reconstruction}

Fig 2. shows a representative sample of the reconstructed coherent
and thermal states. We reconstruct a series of coherent and thermal
states, using the algorithm given by Eq. (1), iterated $10^{6}$ times.
For the quality of our reconstruction we use the fidelity, defined
as $F=\sum_{n=0}^{n=30}\sqrt{\rho_{nn}\widetilde{\rho}_{nn}}$, where
$\tilde{\rho}$ represents the density matrix of the coherent state
corresponding to the average number of photons found in the reconstruction.

\begin{figure}
\includegraphics[width=9cm]{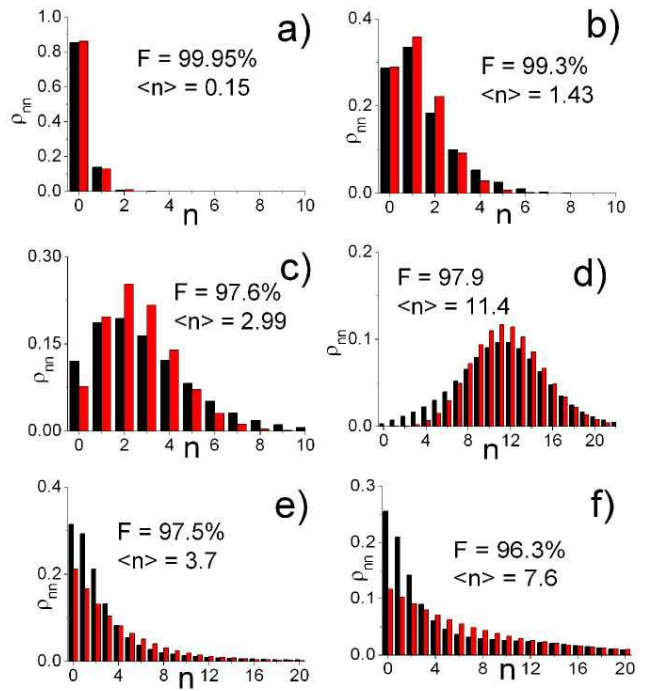}

\caption{Six typical reconstructed states. a)-d) show coherent states, e) and
f) are thermal states. The red bars show the closest coherent (for
a-d) or thermal (e \& f) state, the black bars show the reconstructed
state, i.e. the result of eq. 1 after $10^{6}$ iterations. The fidelities
and mean photon numbers are indicated for each reconstructed state}
\end{figure}

Fig. 3 shows the fidelity of the state reconstruction, as a function
of mean photon number$\langle n\rangle$. We observe that the quality
of the reconstruction degrades as the average number of photons increases.
This can be understood from Fig. 1: as the number of photons increases,
the curves lie closer together, making it more difficult to distinguish
the contributions from various photon numbers.

The theoretical curves in Fig. 3 were generated by simulating the
experiment. Each experiment was simulated 30 times, to obtain a reasonable
estimate of the expected fidelity. The simulations were performed
by calculating expected count rates from the POVM and a given state.
For each calculated count rate curve we assumed a constant relative
error. We made the approximation that these errors are uniformly distrubuted
within the interval $[R-\Delta R,\: R+\Delta R]$.

We compared the mean square error $\chi^{2}$ for the reconstructed
coherent and thermal states with the theoretically expected count
rates for coherent and thermal statistics. From this analysis we conclude
that we can successfully distinguish between coherent and thermal
states. At larger values of $\langle n\rangle$ the value of $\chi^{2}$
becomes large, indicating that the state reconstruction becomes inaccurate
and loses its ability to correctly predict the quantum state. This
happens at $\langle n\rangle\approx9$ and $\langle n\rangle\approx15$
for the thermal and coherent states respectively.

By comparison with the measurements, we find that the relative error
$\Delta R/R$ is 2\% for the coherent states, and 6\% for the thermal
states. These numbers are justified by the observed deviations between
count rates expected from the reconstructed states and the measured
count rates. We attribute this error to the uncertainty in setting
the bias current through the device, where $\Delta R/R=2\%$ corresponds
to 40 nA of uncertainty in the bias current. We attribute the higher
uncertainty for the thermal states to residual variations in the input
intensity caused by the rotating ground glass plate.

\begin{figure}
\includegraphics[width=10cm]{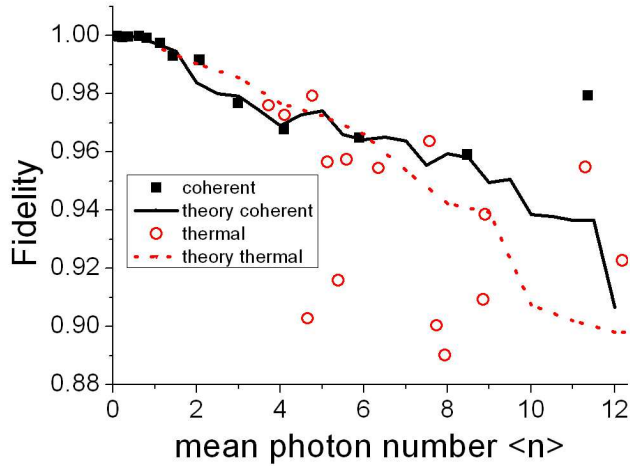}

\caption{Fidelity of the reconstructed states as function of mean photon number.
The solid black squares indicate reconstructed coherent states, the
open red circles indicate reconstructed thermal states. The lines
indicate the results for the expected fidelities from simulations.}
\end{figure}

\section{Nonlinearity-enhanced QSR}

In Fig. 4, we show the effect of the Poissonian (shot-noise) errors
on our reconstruction, calculated from the Cramér-Rao bound \citep{Rossi2004a,Zambra2005}.
In this figure, we compare the expected error bars for state reconstruction
of our SSPD with those of an APD of linear efficiency equal to the
SSPD. The shot-noise error is the fundamental lower limit on the error
and is fixed for a given state. Therefore, it is relevant to investigate
the state reconstruction abilities of a detector.

Fig. 4 shows that the errors in the reconstruction of the SSPD are
50\% lower than those of an APD with equivalent efficiency. We attribute
this to the nonlinear effects which give our detector higher efficiency
at the multiphoton level \citep{Akhlaghi2009}. The physical reason
for this nonlinearity is that two photons which are absorbed close
together on the nanowire have an enhanced probability to make the
detector click. This is in contrast with an avalanche photodiode,
where - as long as each photon overcomes the bandgap - there is no
mechanism where the photons assist one another in producing an avalanche.

\begin{figure}
\includegraphics[scale=0.9]{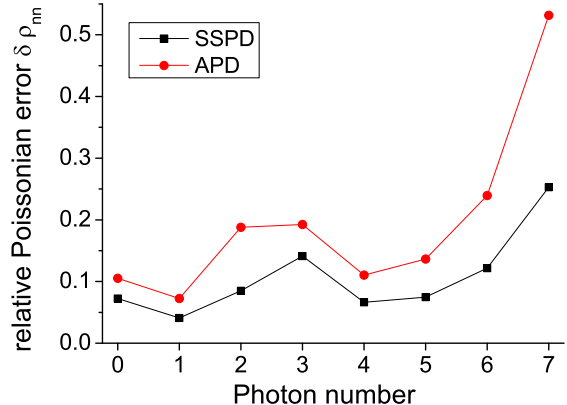}

\caption{Theoretical limitations on state reconstruction, as calculated from
the Cramér-Rao bound, for the same coherent state ($\langle n\rangle=2.5)$
measured by two different detectors. This figure shows the limits
on the ability of the detector to find the components $\rho_{nn}$
of this state. We show the relative error, i.e. $\delta\,\rho_{nn}=\sigma_{\rho_{nn}}/\rho_{nn}$.
The black points indicate the errors expected for our SSPD, the red
ones are for an APD of linear efficiency equal to our SSPD, where
the linear efficiency is the tuning parameter. The errors are calculated
assuming $6*10^{8}$ measurements divided over 100 settings of the
tuning parameter, corresponding to 30 s experimentation time. This
figure shows that the nonlinear effects in the SSPD are beneficial
for state reconstruction. }
\end{figure}

\section{Discussion}

In previous state reconstruction experiments, avalanche photodiodes
(APDs) were used as detectors \citep{Zambra2005}, where the tuning
parameter was the attenuation of an extra attenuator inserted in front
of the detector. These detectors have a POVM element determined only
by linear attenuation, combined with single-photon threshold behaviour.
Banaszek has noted before \citep{Banaszek1998} that state reconstruction
is not limited to $\Pi_{\nu n}$ containing only linear attenuation.
To our knowledge this is the first time that state reconstruction
has been performed with a POVM that contains nonlinear as well as
linear terms.

We find that the practical limitations of our experiment are the accuracy
with which we can set the current. We note, however, that this is
no fundamental limitation, as there are current meters available which
have a much higher resolution than the one used in our experiment.
Furthermore, we note that the Cramer-Rao bound (CRB) has the usual
square-root dependence on the number of measurements made, indicating
that longer measurement times will improve the quality of the reconstruction
as well.

The EM algorithm is known to converge to a solution that saturates
the CRB, meaning that we achieve the lowest possible variance (i.e.
the optimal solution) for our reconstruction. Moreover, through propagation
of errors, the limits on state reconstruction set hard limits on the
ability of the detector to measure other quantitites which are functions
of the input state, such as the second-order correlation function
$g^{(2)}(0)$ \citep{VanderVaart2000}. Therefore, state reconstruction
is a fundamental tool to investigate detectors operating at the few-photon
level, e.g. multi-element SSPD detectors\citep{Marsili2009a,Marsili2009,Jahanmirinejad,Divochiy2008}.
Since state reconstruction is performed at the limit of the amount
of information that can be extracted from a measurement, investigating
the state reconstruction abilities of a detector may yield better
understanding of its capabilities. Furthermore, since QSR probes the
limits of the abilities of a detector, we propose it as a useful benchmark
tool to compare various SSPDs.

In this work, we have used a commercial SSPD, with an efficiency of
up to 2.8\%. However, recently, great strides have been taken in making
SSPDs more efficient, by including cavity structures on the SSPD \citep{Rosfjord2006,Eduard,Kermanproc},
by switching to different materials such as WSi \citep{Marsili93}
and by incorporation of SSPDs in nanophotonic waveguides \citep{Goltsmanarxiv,Dondu}.
Marsili \emph{et al} report efficiencies as high as 93\%. The present
work opens up the possibility of using such high-efficiency devices
for quantum state reconstruction.

It is an open question whether these more efficient detectors would
have beneficial multiphoton nonlinearities similar to the detector
reported on in the present work. We note that this nonlinear effect
has also been reported in \citep{Akhlaghi2009}, and since it is due
to a physical effect that will also be present in more efficient detectors
(namely the absorption of two photons close together) there is nothing
that precludes it. Detector tomography of a high-efficiency SSPD is
necessary to answer this question.

The present work opens up the possibility of applying the ideas of
optimal experimental design to the design of SSPDs. For quantum tomography
of a spin-1/2 system, there has been work \citep{Nunn2010} on how
to optimize the POVM used in a measurement to yield the optimal tomography
result. Similar reasoning may be applied to design an SSPD that is
particularly suitable for reconstructing particular states, or for
measuring particular properties of states. Such optimization - constrained
by what is possible with present production techniques - would focus
on the width of the wire (which governs the nonlinearity), length
of wire segments, number of elements in a multi-element detector,
and the relative size and detection efficiency of each element.

\section{Conclusions}

We have shown quantum state reconstruction with an SSPD. The SSPD
is especially suited for this task because of its low noise, fast
response time, its intrinsic tuning parameter in the form of the bias
current through the device, and the nonlinearities which enhances
the state reconstruction abilities.

Since the fundamental physics of this device is not known, we have
performed a detector tomography procedure in order to find the parameters
describing the response of this device (POVM). We thus demonstrate
state reconstruction with an experimentally determined POVM. This
illustrates the utility of detector tomography.
\begin{acknowledgments}
We thank T. B. Hoang for experimental assistance with the detector.
This work is part of the research programme of the Foundation for
Fundamental Research on Matter (FOM), which is financially supported
by the Netherlands Organisation for Scientific Research (NWO) and
was partially funded by the European Commission through FP7 project
Q-ESSENCE (Contract No. 248095).\bibliographystyle{plainnat}
\bibliography{statreqbib}
\end{acknowledgments}

\end{document}